\documentclass[useAMS,usenatbib]{mn2e}
\usepackage{graphicx}
\title[Astronomical site selection: On the use of satellite data for aerosol content monitoring]{Astronomical 
site selection: On the use of satellite data for aerosol content 
monitoring}
\author[Antonia M. Varela]{A.M. Varela$^{1}$ \thanks{E-mail:avp@iac.es}, C. Bertolin$^{2,3}$, C.Mu\~noz-Tu\~n\'on$^{1}$, 
S. Ortolani$^{3}$ and J.J. Fuensalida$^{1}$ \\
$^{1}$Instituto de Astrof\'{\i}sica de Canarias, Spain\\
$^{2}$National Research Council (CNR), Institute of Atmospheric Sciences and Climate, Padova, Italy\\
$^{3}$Department of Astronomy, University of Padova, Italy \\}

\begin{document}

\date{Accepted 2008 August 5 Received 2008 August 1; in original form 2008 January 10.}

\pagerange{\pageref{firstpage}--\pageref{lastpage}} \pubyear{2008}

\maketitle

\label{firstpage}

\begin{abstract}

The main goal of this work is the analysis of new approaches  to the study of the properties of astronomical sites. 
In particular, satellite data measuring aerosols have recently been proposed as a useful technique for site 
characterization and
searching for new sites to host future very large telescopes. Nevertheless, these
data need to be critically considered and interpreted in accordance with the 
spatial resolution and  spectroscopic channels used. In this paper we have explored and retrieved measurements 
from satellites with high spatial and temporal resolutions  
and concentrated on channels of astronomical interest. The selected datasets are OMI on board the NASA  Aura satellite and MODIS on board the 
NASA Terra and Aqua satellites. A comparison of remote sensing and {\it in situ} techniques is discussed.
As a result, we find that aerosol data provided by satellites up to now are not reliable enough for aerosol 
site characterization, and {\it in situ} data 
are required.

\end{abstract}


\section{Introduction}

Most aerosols reaching the Canary Islands are either marine, ClNa, cryogenic emissions or of African (Sahara and Sahel) origin The latter 
(clays, quartzes, feldspars and calcites), because of their size, can reduce visibility in the optical 
wavelength range and can therefore affect astronomical observations.

Furthermore, aerosols cause radiative forcing, oceanographic deposits by winds (together with Fe and Al),
nutrients and minerals for algae (a coastal increase in 
chlorophyll---phytoplankton biomass), 
sanitary effects, etc. Aerosols also play an important role in astronomical site conditions, 
producing more stable condensation nuclei, delaying precipitation and causing the extinction, absorption, 
diffusion and reflection of extraterrestrial radiation.

Most of the airmass flux component reaching the Canarian archipelago comes from the  North Atlantic Ocean 
 and consists of sea aerosols, which absorbs chloride in the UV. African dust intrusions affect the western and eastern Canary Islands differently.
 Moreover, the presence of a stable inversion layer and the pronounced orography of the western islands 
(La Palma and Tenerife) produce different mass  flux patterns in the lower (mixing) layers closer to the sea  (up to 800 m) and 
in the median-upper (or free) troposphere layer  (above the thermal inversion layer, i.e. above 1500 m), causing a seasonally dependent 
vertical drainage of airborne 
particles.

There are remarkable differences  between summer intrusions, which can rise to the peaks of the
mountains (high level gloom), at 2400 m, and those of winter, more frequent in the lower troposphere (anticyclonic gloom).

Anticyclonic gloom is associated with strong, stationary anticyclonic conditions forced by dust accumulation
between the soil and the inversion layer. It can favour the decrease
in height of the inversion layer and the formation of clouds that do not easily 
precipate as rain, hence 
persisting for a longer time and providing a more stable sea of clouds.
The dust is trapped  by the sea of clouds (sea of dust) and is prevented from reaching the topmost level 
in the islands (above 1500 m).

The aerosol index provided by the TOMS (Total Ozone Mapping Spectrometer) is one of the most widely accepted 
products for detecting the daily  aerosol content. TOMS Level 3 data are gridded in squares of 
 1$^{\circ}\times 1.25^{\circ}$ (latitude and longitude respectively)
and are available online at 
http://toms.gsfc.nasa.gov. The spatial coverage is global 90$^{\circ}$S-90$^{\circ}$N and the temporal resolution is daily.
Moreover, several techniques have been developed {\it in situ} 
to characterize the presence of dust locally at the Canarian  observatories. In particular, a parameter 
related to sky transparency, the atmospheric extinction coefficient in the $V$ (551 nm) and 
$r'$ (625 nm) bands has been measured at the Observatorio del 
Roque de los Muchachos (ORM) on La Palma since 1984  by the Carslberg Automatic Meridian Circle 
(CAMC). The archive is in the public domain at http://www.ast.cam.ac.uk/$\sim$dwe/SRF\-/camc\_extinction.html and provides a good 
temporal comparison with 
the values retrieved with remote sensing techniques from the Total Ozone Mapping Spectrometer on board the Nimbus7 satellite 
and from other probes (Aura/OMI, Terra/MODIS and Aqua/MODIS, MSG1(Met8)/SEVIRI and ENVISAT/SCIAMACHY).
Our main aim is to overlap  the geographical area of the ORM with the satellite data; 
for this reason we have used Level 2 data. Level 0 data are the raw'' data from the satellite; 
Level 1 data are calibrated and geolocated, keeping the original sampling pattern;
the Level 2 data used in this paper are converted into geophysical parameters but still with the original sampling pattern;
finally the 
Level 3 data are resampled, averaged over space, and interpolated/averaged over time
(from http://people.cs.uchicago.edu/$\sim$yongzh/papers/\-CM\_In\_Lg\_Scale\_Production.doc).

On examination, the Level 2 data have the same spatial resolution as the Instantaneous Field of View (IFOV) 
of the satellite. Through a software procedure, it is possible to create files containing 
information on geophysical variables (data describing the solid earth, marine, atmosphere, etc., properties 
over a particular geographical area)
and field values such as seconds, latitude, longitude, reflectivity in different channels, 
the ozone column, the aerosol index, 
aerosol optical depth, cloud land and ocean fraction, SO$_2$ and radiance. From remote sensing and {\it in situ} data it is possible 
to trace back to the cloud coverage and climatic trend. 

The purpose of this study is the analysis of new approaches to the study of the aerosol content 
above astronomical sites. 
Our objective is to calibrate the extinction values in the $V$ band (550 nm) (more details in section 3.1)
with remote sensing data retrieved from 
satellite platforms.
  
The paper is organized as follows: Section 2 describes the problem and background; Sections 3 and 4 
concern the comparison of \textit{in
situ} atmospheric extinction data with the aerosol index provided by TOMS (Level 3 data); 
Sections 5 and  6 deal with the
analysis of Level 2 data from other satellites and their validity for site characterization by comparing the
satellite results with {\it in situ} measurements; and the summary and outlook  
are given in Section 7. 

Two appendices have been included with complementary information. Appendix I describes the format 
of the satellite data, indicating the official websites for data access, and 
Appendix II
includes a list of acronyms to aid in following the terminology used in the paper.

\section{Meteorological and geophysical scenarios}

\subsection{The trade wind inversion as a determining factor of aerosol distribution}

Site testing campaigns are at present performed within the classical scheme of  optical seeing properties, meteorological 
parameters, air transparency, sky darkness, cloudiness, etc.  New concepts related to geophysical properties 
(seismicity and microseismicity), local climate variability, atmospheric conditions related to the optical turbulence 
(tropospheric and ground wind regimes) and aerosol presence have recently been  introduced in the era of selecting the
best sites for hosting a new generation of extremely large telescopes (which feature a filled aperture collector larger than 40 m, and which are 
considered worldwide as one of the highest 
priorities in ground-based astronomy),  telescope and dome designs,
 and for feasibility
studies of adaptive optics (Mu\~noz-Tu\~n\'on  2002; Mu\~noz-Tu\~n\'on et al. 2004; Varela et al. 2002; Varela et al. 2006; 
Mu\~noz-Tu\~n\'on et al. 2007). 

The Canarian Observatories are among the top sites for astronomical observations and have been monitored and 
characterized over several decades (Vernin et al. 2002). The 
trade wind scenario and 
the cold oceanic stream, in combination with the local orography, play an important role in the retention of low cloud 
layers well below the summits to the windward (north) side of the islands, above which the air is dry and 
stable (the cloud layers also trap a great deal of light pollution and aerosols from the lower troposphere).

\begin{figure}
\includegraphics[scale=0.65]{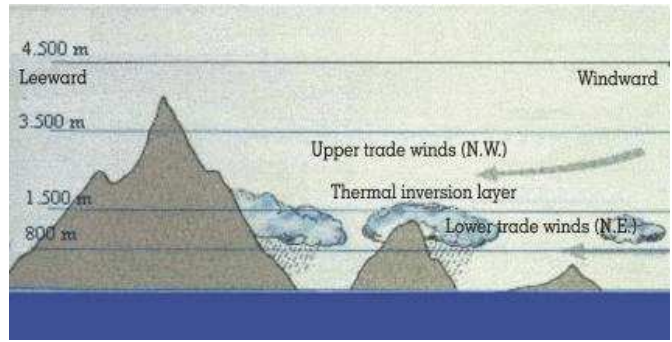}

\caption{Trade wind behaviour in the low mixing maritime  layer (MML) and in the tropospheric 
layer (TL). From http://cip.es/personales/oaa/nubes/nubes.htm.}

\label{fig_trade}
\end{figure}

The trade winds and the thermal inversion layer between 1000 and 1500 m (shown in Fig.\ \ref{fig_trade}) have been 
the object of many studies over the last 50 years, either indirectly from observations of the stratocumulus layer 
(known locally as the ``sea of clouds''), which forms at the condensation level (Font-Tullot 1956) or from radiosondes 
(Huetz-de-Lemps 1969). For much of the year, the trade wind inversion layer separates two very different airmasses: 
the maritime mixing layer (MML) and the free troposphere layer (TL) (Torres et al.\ 2001). 
The presence of the inversion layer is crucial in the airmass flows reaching the islands and in the downflow at high 
elevations that they undergo. Saharan dust invasions mostly affect the eastern islands, but can occasionally reach the 
islands of 
Tenerife and La Palma; normally, however, they do not reach the level of the Observatories (the ORM is 2396 m and 
the OT 2390 m 
above mean sea level).

\subsection{Airmass intrusions in the Canary Islands}

Airmasses are classified into three types according to their origin and permanence over continental 
landmasses and sea: 
Atlantic (marine, ClNa), European (anthropogenic emissions from sulphates and carbon) and African (dust, mineral 
aerosols). The most frequent are chlorides, i.e.\ clean salts of oceanic origin that do not affect 
astronomical observations, and 
that reach or exceed half the total contribution of mass flux over the summits. 
European airmasses are almost always 
of anthropogenic origin (sulphates and carbon) and are of scant importance (between 0.8 and 7.2\%; see Romero 2003).

A  study of the origins and distribution of air flows performed by Torres et al.\ (2003) distinguished  
the origins of 
airmasses that affect the lower  (MML) and upper (TL) layers, the latter being at the height 
of the astronomical observatories. 
Daily isentropic retrotrajectories (at 00:00 and 12:00 GMT) were used during the AEROCE  
(Atmosphere/Ocean Chemisty Experiment) at Iza\~na (2367 m above MSL, 1986--97) and the Punta del 
Hidalgo Lighthouse (sea level, 1988--97), each airmass being assigned an origin: North American, North Atlantic, 
Subtropical Atlantic,
Europe, Local and Africa, the summer and winter periods being separated.

The airmass provenance in the lower (mixing) layer close the sea (MML) is Northern Atlantic (the mean is 59.6\%), European  (the
mean is 19\%) 
and African (0\% in summer and  23\% in winter), whereas the troposphere layer (TL) is mainly Northern Atlantic (44.2\%) and 
African (17.4\%), with a minimum in April (5.3\%)  and a maximum in August (34.5\%) (Torres et al.\ 2003).

African summer intrusions are therefore almost absent in the MML and more intense in the TL because 
of the daily thermal convection reaching the higher atmospheric
layers.  In winter intrusions into the TL are less frequent. The airmass is carried horizontally by the 
prevailing wind and  
is affected by a process 
of separation, the larger particles ($>$ 10 $\mu$m) leaving sediments at ground level over a short time and the 
smaller ones being carried across the Atlantic Ocean to distances of hundreds or thousands of kilometres from their 
place of origin.
 This sand creates a large feature (plume) that is visible in satellite images (see, for example, the EP/TOMS 
or Aura/OMI websites)
 and extends from the African coast across a band about 20$^{\circ}$ in latitude. During winter, 
 the prevailing 
 wind carries the dust from the south of Canary islands through an average  of 10$^{\circ}$ in latitude to the Cape 
Verde  Islands.
Considering that the dust plumes reach an extension of about 2000 km, during the winter they rarely reach
La Palma. Instead, in summer, winds above 4 km in altitude can take these particles as far north as
30$^{\circ}$ in latitude. 
In these conditions the dust plume over the ORM is composed mainly of small quartz particles in the range 
0,5--10 $\mu$m and 
the biggest particles precipitate (dry deposition). 
Typical dust storms take 3--8 days to disperse and deposit 
1--2.4 million tonnes/year. Finally, the clouds of aerosols dissipate through advective processes or through rain 
(sumid deposition).

\section{Database for the analysis: {\it in situ} extinction coefficient values and aerosol indexes provided by the 
TOMS and OMI spectrographs on board satellites}
\subsection{Atmospheric extinction coefficient measured above the ORM (La Palma)}

Atmospheric extinction is the astronomical parameter that determines transparency of the sky. Extinction  is associated 
with the 
absorption/scattering of incoming photons by the Earth's atmosphere and is characterized by the extinction coefficient, $K$. 
Sources of sky 
transparency degradation are clouds (water vapour) and aerosols (dust particles included). This coefficient is 
wavelength-dependent and can be determined by making observations of a star at different airmasses. For details of the astronomical 
technique 
for deriving the extinction coefficient values we refer the reader to King (1985). 

Long baseline extinction values for the ORM have been measured continously at the Carlsberg Automatic Meridian Circle 
(CAMC, http://www.ast.cam.ac.uk/\~dwe/SRF\-/camc\_extinction.\\html) in the $V$ band (550 nm)  and more recently in the Sloan $r'$ band (625 nm).
To our knowledge, this is the largest available homogeneous 
database for an
observing site.

Extinction values and their stability throughout the night are essential for determining the accuracy of astronomical 
measurements. As nights with low and constant extinction are classified as photometric, this parameter is considered  
among those relevant for characterizing an observing site.

On photometric dust-free nights the median of the extinction is 0.19 mag/airmass at 480 nm, 0.09 mag/airmass at 625 nm 
and 0.05 mag/airmass at 767 nm. The extinction coefficients reveal that on clear days (denoted coronal-pure) the extinction 
values at 680 nm are below  about 0.07--0.09 mag/airmass, while on dusty days (diffuse-absorbing) they 
are always higher (Jim\'enez \& Gonz\'alez Jorge 1998). The threshold that identifies the presence of dust is at 0.153 mag/airmass 
(Guerrero et al. 1998).

 At the ORM, the extinction in $V$ is less than 0.2 mag/airmass on 88\% of the nights, and extinction in excess of 
0.5 mag/airmass 
 occurs on less than 1\% of nights. A statistical seasonal difference is also detected (see Guerrero et al. 1998).

\begin{figure*}
\includegraphics[scale=0.65]{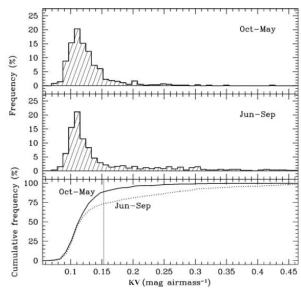}

\caption{Frequency of extinction over the ORM during winter (top) and summer (centre). In both cases the modal value is 
0.11 mag/airmass. Their corresponding cumulative frequencies are also shown (bottom). The vertical line indicates the 
extinction coefficient limit for dusty nights, KV $\geq$ 0.153 mag/airmass (Guerrero et al. 1998).}
\label{fig_stsext}
\end{figure*}

 Figure \ref{fig_stsext} shows the cumulative frequency of extinction over the ORM during winter (October--May) and summer 
(June--September). 
 In summer, 75\% of the nights are free of dust, while at other times of the year over 90\% of the nights are dust-free 
 (Guerrero et al. 1998).  These results are consistent with those provided by Torres et al. 2003.

\subsection{Characterization of aerosols in the near-UV}

Data  from Total Ozone Monitoring Spectrometer (TOMS) and from the Ozone Monitoring Instrument (OMI) are analysed to detect 
absorbing and non-absorbing aerosols at ultraviolet (UV) wavelengths.
A detailed description of these instruments and the products are given at the official Website (http://toms.gsfc.nasa.gov/).
Absorbing aerosols include smoke deriving from biomass burning, industrial activity, mineral dust, volcanic aerosols and soot. 
Non-absorbing aerosols are mostly sulphates.
The UV spectral contrast is useful for retrieving values over land and ocean because of its low reflectivity in 
the spectrometer range. Backscattered radiation at $\lambda$ 340, 360 and 380 nm is  caused mainly by molecular Rayleigh scattering, 
 terrestrial reflection and   diffusion by aerosols and clouds (through Mie scattering).
Quantitatively,  aerosol detection from TOMS and OMI is given by:

\begin{displaymath}
\Delta N_{\lambda}=-100\left\{{\log_{10}\left[{\left({\frac{I_{331}}{I_{360}}}\right)_{\rm meas}}\right]-
\log_{10}\left[{\left({\frac{I_{331}}{I_{360}}}\right)_{\rm calc} }\right]}  \right\},
\end{displaymath}

\noindent where $I_{\rm meas}$ is the backscattered radiance at the wavelenghts measured by TOMS and OMI (with Mie and Rayleigh 
scattering, and absorption) and $I_{\rm calc}$  is the radiance calculated from the model  molecular atmosphere with  Rayleigh 
scatterers.
$\frac{I_{331}}{I_{360}}$  depends strongly on the absorbing optical thickness of the Mie scatterers. 
  $\Delta N_\lambda$ is also called the aerosol index (AI). AI $>$ 0 indicates the presence of absorbing aerosols and clouds
  (AI$\pm $0.2) and negative AI values indicate non-absorbing aerosols (Herman et al. 1997; Torres et al. 
  1998).
  Significant absorption has been set at AI $>0.7$ by Siher et al. 2004 and at AI $>0.6$ in this work (see
  explanation in next section and in Fig. \ref{fig_limits}).

The next section (Results I)
concerns the first comparison of the {\it in
situ} atmospheric extinction coefficient with the aerosol index provided by TOMS using Level 3 data. In 
Section 5  we present recent instruments on board satellites suitable for aerosol content monitoring and in Section 6 (Results II) we 
compare 
the atmospheric extinction coefficient with the aerosol index and aerosol optical depth provided by OMI and MODIS respectively, 
using Level 2 data.

\section{Results I: comparison of AI provided by TOMS with KV from the CAMC}

Atmospheric extinction is related to the internationally recognized term aerosol optical depth (AOD) (or thickness) and to the aerosol index. 
In this section we shall analyse the first results of comparing the atmospheric extinction coefficient with the aerosol index 
 provided by TOMS (Total Ozone Mapping Spectrograph) on board NASA's Earth Probe  satellite. We use Level 3 aerosol data,
which are available at:

ftp://toms.gsfc.nasa.gov/pub/eptoms/data/aerosol/.

In a previous paper
(Varela et al. 2004a) we  presented the aerosol index  from the sector corresponding to the Teide Observatory (Tenerife)
against the atmospheric extinction coefficient recorded at Roque de los Muchachos Observatory (La Palma). 
The Teide Observatory (OT) is situated 2390 m above sea level in Iza\~na and the geographical coordinates are 16$^\circ$ 30$'$ 35$"$ West
and 28$^\circ$ 18$'$ 00$"$ North. 
The Observatorio del Roque de los Muchachos is situated on the edge of the Caldera de Taburiente National Park, 
2396 m above sea level and the geographical coordinates are 
17$^\circ$ 52$'$ 34$"$ West and  28$^\circ$ 45$'$ 34$"$ North. 
The two observatories are about 133 km apart. We   compared the AI  from 
EP/TOMS data centred at the OT and at the ORM sectors (see figure 5 of Varela et al. 2004b).

\begin{figure*}
\includegraphics[scale=0.5]{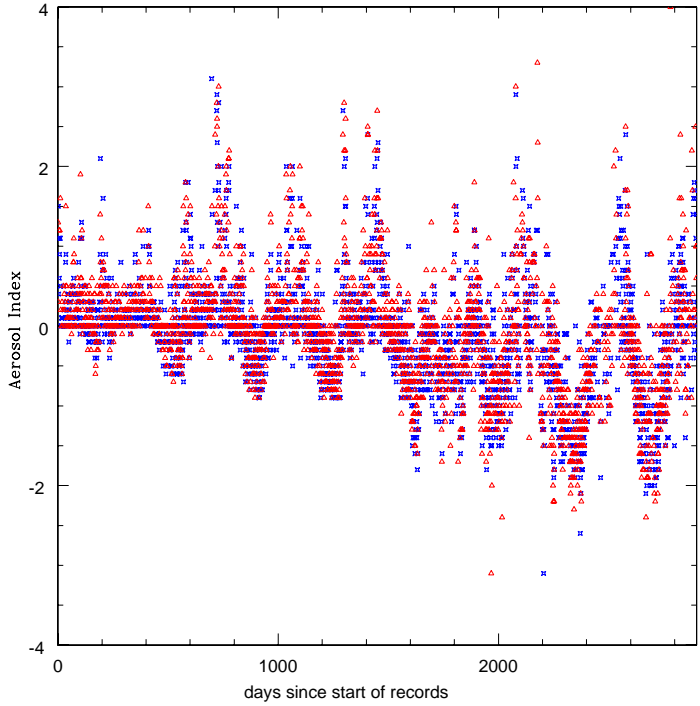}

\caption{Aerosol index provided by EP/TOMS from 1996 to 2004 at the OT (red triangles) and at the ORM (blue crosses). 
These profiles indicate similar aerosol tropospheric distributions at both observatories.}

\label{fig_ORMvsOT}
\end{figure*}

The consistency of AI from both boxes centred on the OT and on the ORM shown in Fig. \ref{fig_ORMvsOT} points to a similar 
tropospheric aerosol distribution at both observatories.

In the next section (Results II) we show the results of comparing the atmospheric extinction coefficient with the aerosol index provided by
OMI on board NASA's  Aura satellite and with the aerosol optical depth provided by MODIS on board NASA's Terra and Aqua satellites.

\begin{figure}
\includegraphics[scale=0.6]{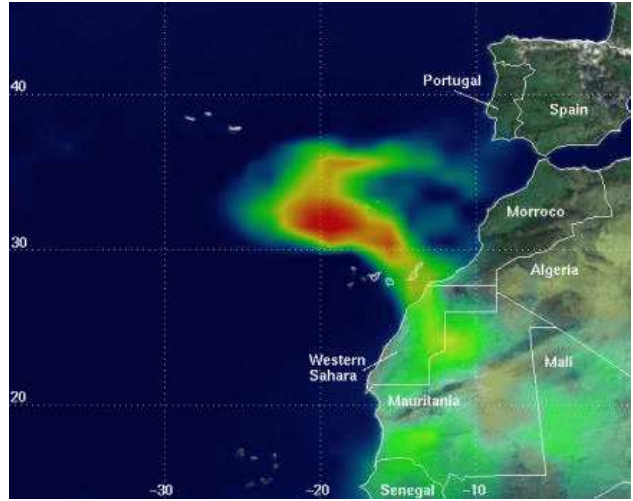}

\caption{Dust plume obtained from TOMS data over the western Sahara Desert and extending over the Atlantic Ocean and Canary islands. 
From http://toms.gsfc.nasa.gov/aerosls/africa/canary.html.}

\label{fig_plume}
\end{figure}

To demonstrate why there is not necessarily any correlation between the AI and atmospheric extinction we 
have selected an intense invasion of African dust over the Canary Islands that occurred on 2000 February 26. Figure \ref{fig_plume}
shows the plume of dust as recorded by TOMS on board the Earth Probe. We have used AI values obtained from TOMS for the Canarian 
Observatories and atmospheric extinction values provided by the CAMC at the ORM during the 
night before and after this 
episode. AI reaches its maximum (2.4 at the OT and 2.7 at the ORM) on February 27, precisely on the day 
when the plume 
reached  Tenerife. Nevertheless, the CAMC measures an extinction value of less than 0.2 mag, with a 
high number of 
photometric hours. The reason is that when the plume arrived at La Palma it did not reach the level of the 
Observatory.

\begin{figure*}
 \includegraphics[scale=1]{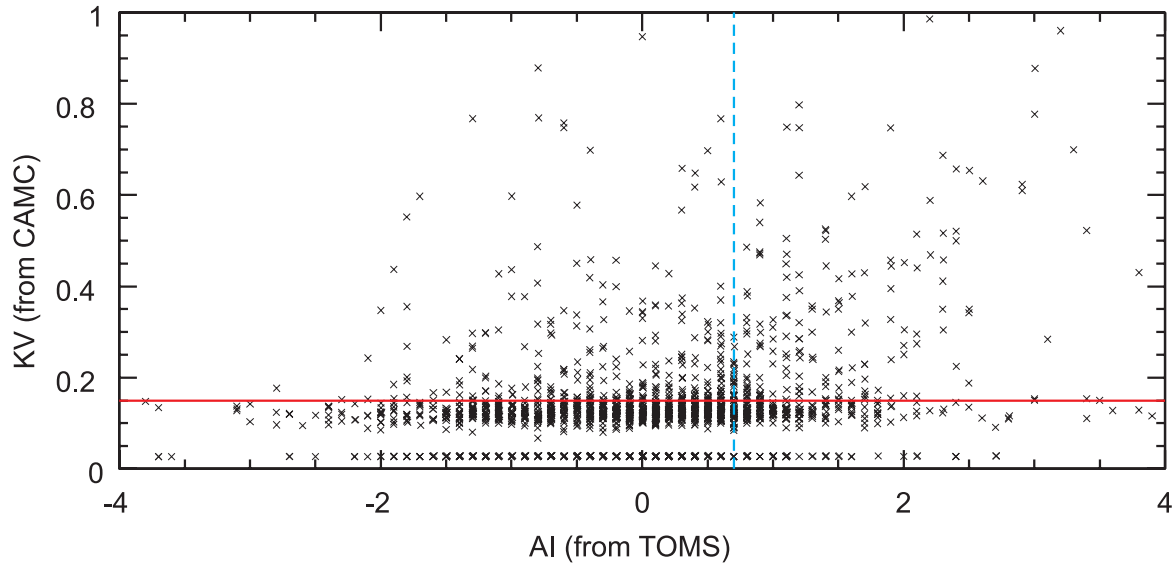}

\caption{Aerosol index provided by TOMS against the atmospheric extinction coefficient in $V$ (integrated values) 
from the CAMC above the  ORM. The red horizontal line is a threshold indicating the presence of dust in the atmosphere 
(Guerrero et
al. 1998). The blue vertical dashed line is the threshold line indicating the presence of absorbing aerosols in the atmosphere. The top-right
quadrant corresponds to points seen as dusty (from CAMC KV $>$ 0.15) and of high absorbing aerosol presence 
(AI $>$ 0.7 has been used by Siher et al. 2004).}

\label{fig_TOMSAI-KV}
\end{figure*}

 In Fig. \ref{fig_TOMSAI-KV} we represent the atmospheric extinction in the V band (KV) provided by the 
 Carlsberg Meridian Circle against the aerosol index provided by 
  EP/TOMS from 1996 to 2004. 

We have classified  four quadrants using KV=0.15 mag/airmass as the threshold for dusty nights and 
AI=0 as the threshold of the presence of absorbing aerosols.

Figure \ref{fig_TOMSAI-KV} shows a large number of TOMS data indicating the presence of absorbing aerosols
coincident with CAMC values that show low or no atmospheric extinction 
 (bottom right   in Fig. \ref{fig_TOMSAI-KV}). This result is due to the presence of a layer of  dust below the Observatory level. This dust layer is high 
 and/or thick enough to be detected by the TOMS. This condition appears in  17\% of all cases.

 The case of agreement between large extinction coefficient and large AI  (top right  in Fig. \ref{fig_TOMSAI-KV}) is associated with dust 
 presence in the upper troposphere layer (TL). We have verified that this case occurs in only 11\% of all measurements, with
 58\% of these corresponding to the summer months (June--September), when the warmest surface winds sweep dust from the African 
 continent and rise towards the upper layers by convective processes (Romero \& Cuevas 2002; Torres et al. 2003).

 The opposite case, i.e. low extinction coefficient and low aerosol index  (bottom left in Fig. \ref{fig_TOMSAI-KV}) happens in 59\% of cases.

 In the top left quadrant in Fig. \ref{fig_TOMSAI-KV}, we find low AI values of aerosol index but large KV  
 (13\% of the data points). A possible explanation is given by  Romero \& Cuevas 2002, who argue 
 that the cause could be local and  concentrated dust in a small area intercepting the light from a star (and therefore measured by CAMC) 
 but that is 
 not representative of the average values from  
 a 1$^{\circ}\times 1.25^{\circ}$  region provided by TOMS;  or is due to the presence of high fine dust layers above the 
 Observatory that are not detected by TOMS. Here we propose  cirrus or other clouds as the explanation; in fact, 78\% 
 of the points located in the top-left quadrant correspond to the winter months when the possibility of the presence of medium-high clouds is greater.

In the four cases (four quadrants), the correlation coefficient and the square of the coefficient of the Pearson  correlation is smaller than 
0.1, so there is no correlation at all between both parameters. The maximum correlation is found in the upper-right 
quadrant (mathematically, this is the first quadrant), 
i.e. AI positive and extinction coefficient larger than 0.15. When this interval is narrowed to AI larger than 0.7 and KV 
larger 
than 0.2 mag/airmass (only  4\% of cases), this correlation increases slightly. Only when the summer period is considered 
under 
these last conditions can the correlation reach 0.55.

 This last result coincides with Siher et al. 2004 (see Fig.2 in this paper), correlating AI TOMS data recorded by Nimbus7 and the atmospheric 
 extinction 
 coefficient at the ORM during summertime dusty events (AI$>$0.7 and KV$>$0.2). The correlation found is 0.76. 
 It is important, however,
  to emphasize that 
 this relatively high (0.76) correlation occurs for a subsample  (4\%) of the total cases that occur when AI$>$0.7 and KV$>$0.2. 
 As result of the lack of statistical meaning in this result, the yearly average extinction derived from TOMS 
 (figure 6 in Siher et al. 2004)
 does not reflect the real values measured {\it in situ} at the Observatory.
 
 This correlation occurs in summertime, when the warmest winds can drive the dust to the level of the
 Observatory.

The non-correlation between AI satellite remote sensing data and in situ extinction data is also reported by 
Romero \& Cuevas 2002 (see figure 3 in this paper),  
who compare the aerosol index (AI) provided by  TOMS  against the  optical aerosol depth (AOD) obtained with a multifilter 
rotating 
shadow band radiometer (MFRSR) at Iza\~na Atmospheric Observatory (OAI) (close to the Teide Observatory, 133 km the ORM), an institution
belonging to the Instituto Nacional de Meteorolog\'{\i}a (INM), now called Agencia Estatal de Meteorolog\'{\i}a (AEMET). The MFRSR measures 
global 
and diffuse radiation in six narrow band channels between 414 nm and 936 nm (with bandwidths between 10 and 12 nm).

Again, in Romero \& Cuevas 2002 a low correlation coefficient is attained when the AOD is larger than 0.1 at 414.2 nm and the AI is larger 
than 
0.5---coinciding with dust invasions taking place owing to convective processes from the African continent towards high levels---and will then 
be detected by both TOMS and the MFRSR. Otherwise, in dust-free conditions, there is no correlation between 
the {\it in situ} measurements (AOD) and AI. 

The prevailing trade winds, the inversion layer presence and the abrupt orography of the western Canary islands (La Palma and Tenerife)
play an important role in the downflow of dust at high elevations.

The low spatial resolution of the EP/TOMS for astronomical site evaluation explains the absence of correlation; this spatial resolution
cannot 
distinguish local effects since it averages over  a surface equivalent to an entire island
(a 1$^{\circ}\times 1.25^{\circ}$ region given for TOMS Level 3 data made available in GSFC TOMS archives, i.e. 111 km $\times$ 139
km); 
neither does it distinguish the vertical dust drainage. This effect is known as anticyclonic gloom, in many cases, even in
such  dust
storms episodes, the airborne dust particles do not necessarily reach the level of the Observatory (more explanations in Section 2 of this
paper).

It is possible to retrieve high resolution data from TOMS instrument also with the spatial resolution of the IFOV 
(35 km $\times$ 35 km) or TOMS Level 2 data (see Bertolin 2007) in order to show that the correlation between CAMC KV and  EP/TOMS AI for 
the ORM site 
on La Palma shows a little improvement when data are analysed at higher spatial resolution.

The next step in our study is to explore the use of other instruments
on board different
satellites that operate in bands of
astronomical interest (the visible and NIR) with higher spatial
resolution than TOMS and with long-term databases (longer than a few
years). These instruments will provide updated high resolution images
for astronomical site evaluation (to date,  only TOMS data have been
used for aerosol content characterization above an astronomical  observatory).
 
\section{Recent instruments on board satellites for aerosol content monitoring for astronomical site evaluation}

\subsection{Updated images}

We have explored  the use of other detectors on board different satellites that operate in bands of 
astronomical interest (the visible and NIR) with higher spatial resolution than TOMS. 

The new generation of satellites  
Terra$\&$Aqua/MODIS, Aura/OMI, MSG1/SEVIRI, etc.---acronyms described in Appendix II---provide better-resolution images 
that can
be used to demonstrate the existence or otherwise of any correlation between the presence of aerosols and atmospheric
extinction. Similarly to the case mentioned above (February 2000), we  examine
2007 March 10,  when a thick plume of dust blew   over the Canary islands from the west coast of Africa.

\begin{figure*}
  \includegraphics{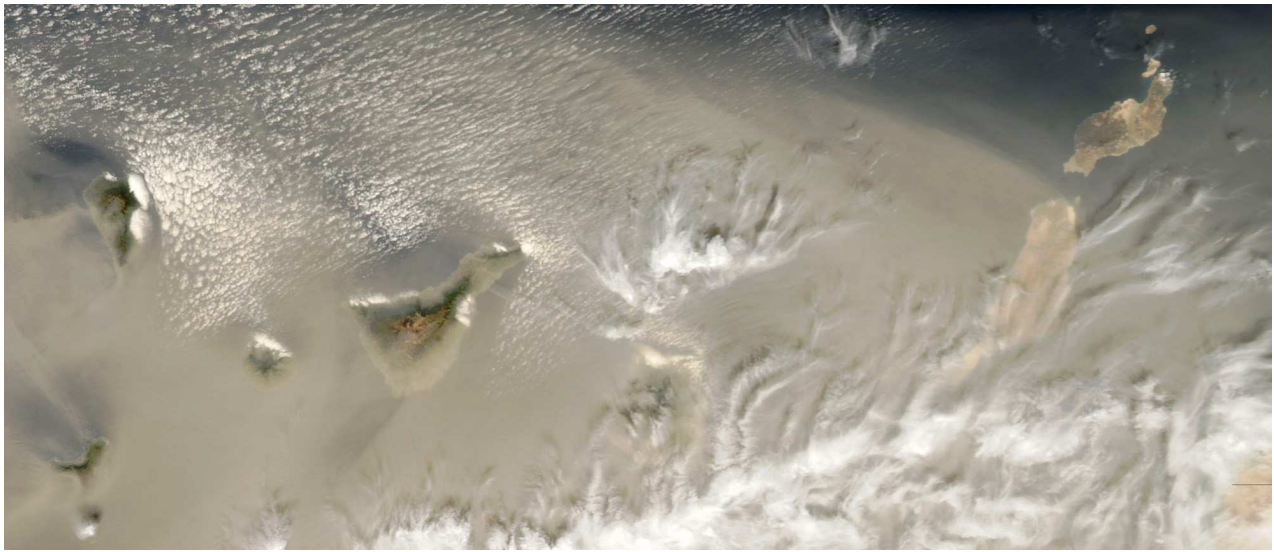}
   \hspace*{\columnsep}%
\caption{Plumes of dust   over the Canary Islands from the west coast of Africa observed by  
the Moderate Resolution Imaging Spectroradiometer (MODIS) on NASA's Aqua satellite on 2007 March 10
(full image at http://earthobservatory.nasa.gov/Newsroom/NewImages/Images/canary$\_$amo$\_$2007069$\_$lrg.jpg). 
 The easternmost island is just over 100 kilometres from the African coast. The peaks of
La Palma and Tenerife remain clear of dust.}
\label{fig_modisplume}
\end{figure*}

The Moderate Resolution Imaging Spectroradiometer (MODIS) on NASA's Aqua satellite took the picture shown 
in Fig. \ref{fig_modisplume} for this day 
(with much better resolution than TOMS,  see Table 1). 	
The MODIS Aerosol (MOD04$\_$L2) product monitors the ambient aerosol optical thickness over the oceans globally and over a portion of the 
continents and contains data that have a spatial resolution (pixel size) of 10 km $\times$ 10 km (at nadir). 
More information about grids and granule coverage is available in the official MODIS 
website (http://modis.gsfc.nasa.gov/). We can see that the eastern islands are more affected
by the dust plume than the western ones. Also, the abrupt orography and the inversion layer play an important role in
retaining
the dust below the summits (where the observatories are located) of the highest western islands (Tenerife and La Palma).
This effect is the above-mentioned anticyclonic gloom. 

This situation has been verified by data provided by the CAMC that indicate no atmospheric extinction at all:
\begin{itemize}
 \item{Extinction Coefficient in r$'$= 0.083 (good quality dust-free night) i.e. 
      KV =  0.12 mag/airmass.}
 \item {Number of hours of photometric data taken=10.22}
 \item {Number of hours of non-photometric data taken=0.00}
 \end{itemize} 

In the next section we summarize data retrieved  from the new generation of  
satellites.
Despite the much better spatial and spectral resolution of the recent satellite aerosol
measurements (AI and AOD), they are not at the moment sufficient for the aerosol content monitoring of an astronomical 
observatory. {\it In situ} 
data are also required, in particular at those astronomical sites with abrupt orography 
(ORM, Mauna Kea or San Pedro M\'artir). Spatial resolution of the order of the observatory area is needed.

\subsection{Updated data}

In order  benefit from satellite data for local site characterization, we have gathered and studied NASA and ESA 
satellite data planned to retrieve information about aerosol, clouds, 
ozone and other trace gases (N$_2$, O$_2$,  H$_2$O, CO$_2$, CH$_4$) that are found in the terrestrial atmosphere. 
In this paper we
have centred our analysis on the aerosol content.  

\begin{figure*}
\includegraphics[scale=1]{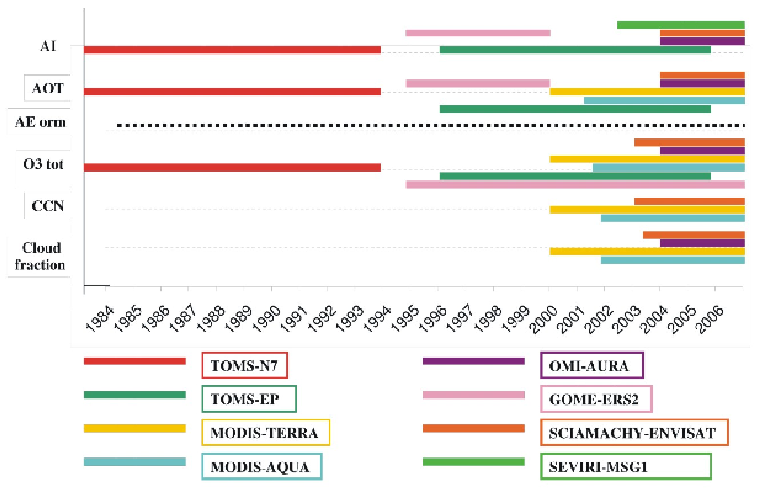}

\caption{Overview of parameters (vertical labels) and periods provided by instruments on board 
ESA and NASA satellites (shown with different colour codes). For the aerosol content only Aura/OMI, Terra/MODIS and Aqua/MODIS results
are shown in this paper to provide  high spatial resolution and a longer-term database. 
Data from {\bf ERS2/GOME and 
MSG1/SEVIRI} have not been  
included in this work because the databases are too short-term for statistical analysis. The {\bf ENVISAT/SCIAMACHY} database has not been used in 
this
analysis because it does not provide
higher spatial resolutions than OMI and MODIS.}
\label{fig_GANT}
\end{figure*}

An overview of selected satellites, parameters and sampled periods is given in Fig. \ref{fig_GANT}, 
which includes other
parameters (ozone, cloud condensei nuclei (CCN) and cloud fraction) to be analysed in a future paper.

To ensure that the retrieved 
remote sensing data fields from different satellites (such as aerosol values or geolocation parameters) are precisely
over the ORM site coordinates to compare  with the atmospheric extinction by CAMC, 
we decided to work with Level 2 data, which have a projected effective pixel 
size given by the instantaneous field of view (IFOV). We have also selected the longer term database (to retrieve more data
for the KV comparison). The satellites/instruments and parameters used in this work are summarized in Table 1.

The previous parameters selected and retrieved are the
aerosol index (AI) provided by OMI (Ozone Monitoring Instrument) on board Aura---with visible and ultraviolet 
channels and with a spatial resolution from 13 km $\times$ 24 km to 
24 km $\times$ 48 km---and the aerosol optical depth (AOD) provided by MODIS (Moderate Resolution Imaging Spectroradiometer) 
on board Terra (from 2000) and Aqua (from 2002)---with its 36 spectral bands, from 0.47 to 14.24~$\mu$m, 
including two new channels 0.405 and 0.550~$\mu$m, 
with a spatial resolution of 10 km $\times$ 10 km. A detailed description of the instruments, 
parameters and data access url is given by Varela et al. 2007. Data from ERS-2/GOME and 
MSG1/SEVIRI   have not been  
included on this work because databases are too short-term for statistical analysis, but they do offer valuable potential as instruments 
for the future when the database increases. The ENVISAT/SCIAMACHY database has not been used in 
this analysis because it does not provide
higher spatial resolutions than OMI and MODIS.

\begin{table*}
\caption{Overview of instruments on board satellites that provide  parameters useful for our work}
\begin {tabular*}{\hsize}{@{\extracolsep{\fill}}lcccc}     \hline
SATELLITE/INSTRUMENT	&HORIZONTAL RESOLUTION	&PARAMETER	&PERIOD\\ \hline
Terra/MODIS&	10$\times$10 km$^2$	&Aerosol Optical Depth (AOD)&	from 2000 \\
Aqua/MODIS&	10$\times$10 km$^2$	&Aerosol Optical Depth (AOD)&	from 2002\\
Aura/OMI&	From 13$\times$24 km$^2$ to 24$\times$48 km$^2$&	Aerosol Index (AI)&	 from 2004\\
\hline 
\end{tabular*} 
\end{table*}

In Appendix I we indicate the official websites to retrieve datasets and the data formats. The correlation analysis between AI and AOD with KV is given in Section 6.

\subsection{Using Aura/OMI data}

Here, we show the results of comparing AI provided by TOMS and OMI. 
OMI on the EOS (Earth Observing Systems) Aura platform 
continues the TOMS record for total ozone and other
atmospheric parameters and can distinguish between aerosol types, such as smoke, dust and sulphates.

\begin{figure*}
\includegraphics[scale=1]{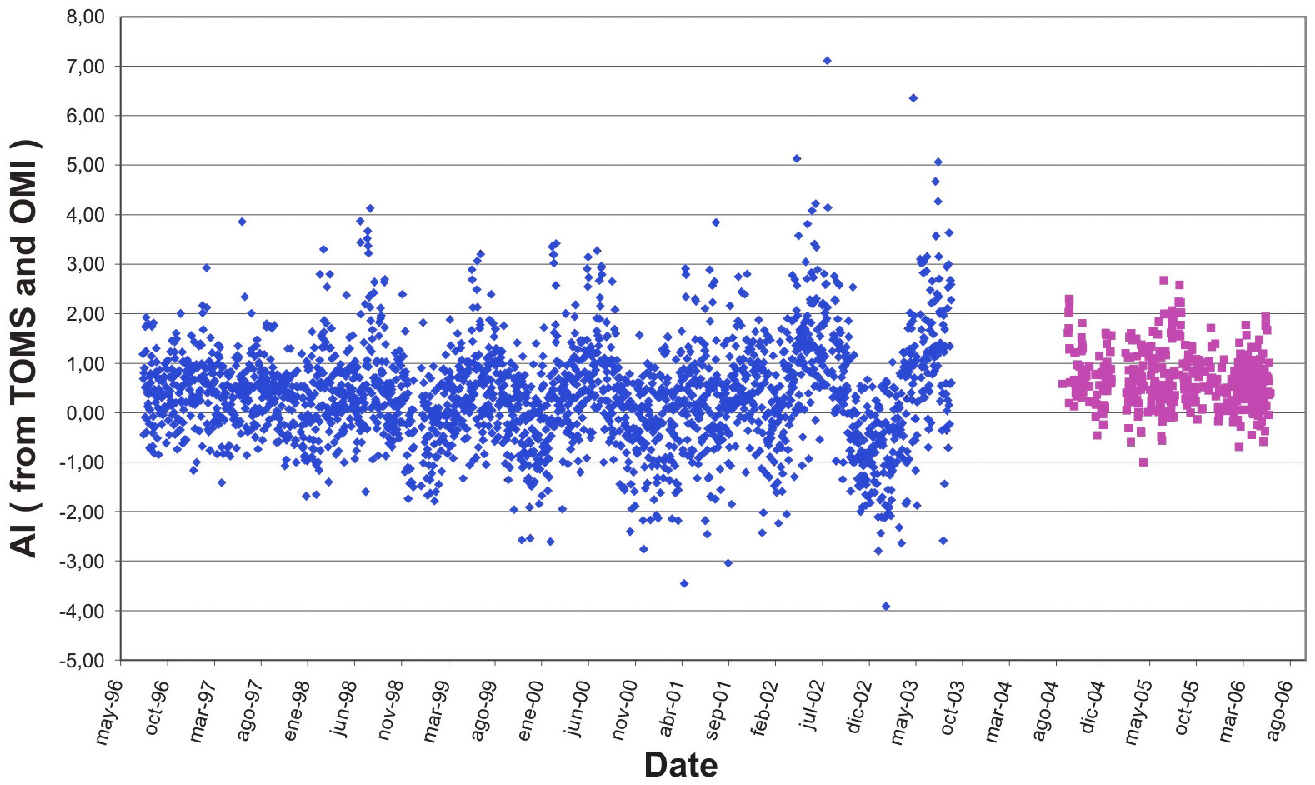}

\caption{Aerosol Index provided by  EP/TOMS (blue points on the left) and  Aura/OMI (pink points on the right) in time 
over the Roque de los Muchachos Observatory on La Palma. The 
improvement
shown in the OMI data (much less dispersion than TOMS data) derives from better horizontal and vertical spatial resolution 
compared with its
predecessor, TOMS in Earth Probe.}
\label{fig_AI}
\end{figure*}

Figure \ref{fig_AI} shows the AI values provided by EP/TOMS and OMI/Aura over the Roque de los Muchachos Observatory on La Palma. Note the dispersion in the
TOMS data. This improvement
shown in the OMI data (much less dispersion) derives from better horizontal and vertical spatial resolution 
compared with its
predecessor, TOMS in Earth Probe. Moreover, TOMS data from 2002 should not be used for trend analysis because of calibration errors
(http://jwocky.gsfc.nasa.gov/nes/news.html).

\subsection{Using Terra/MODIS and Aqua/MODIS data}

MODIS on board the NASA Terra and Aqua satellites provides not AI values but an
equivalent parameter, the aerosol optical depth (AOD).

\begin{figure*}
\includegraphics[scale=1]{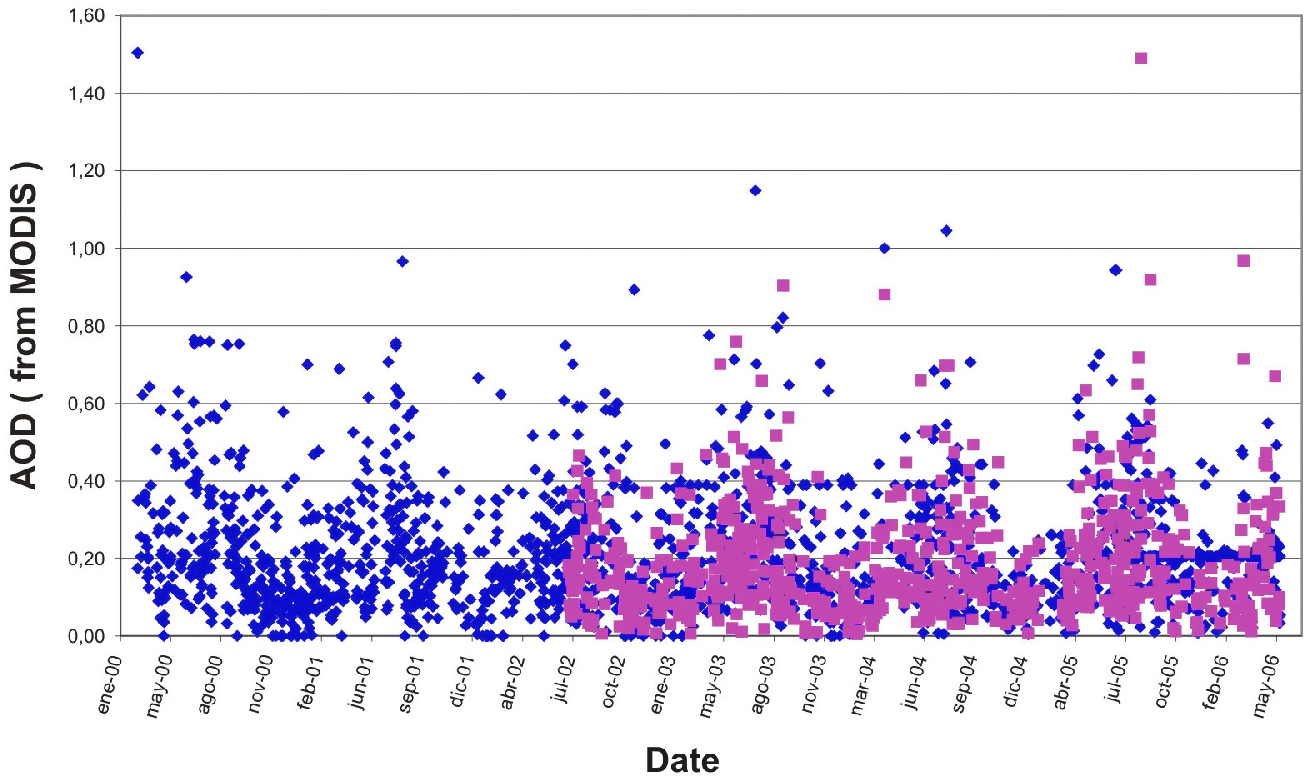}
\caption{Aerosol optical thickness provided by  Terra/MODIS (blue points) and Aqua/MODIS (pink points) in time over the Roque de los
Muchachos Observatory on La Palma.}
\label{fig_AOD}
\end{figure*}

In Fig. \ref{fig_AOD} we show AOD for Terra and Aqua; we see that there exists a good relation between the maximum 
and minimum AOD values. The consistency of both data sets is excellent. 

The AI threshold for dusty nights is 0.6; we now have to determine the AOD threshold for dusty events, but 
AI and AOD are not in a 1:1 ratio (they depend on refractive index, particle
size distribution and height of the atmospheric layer).

\begin{figure*}
\includegraphics[scale=1]{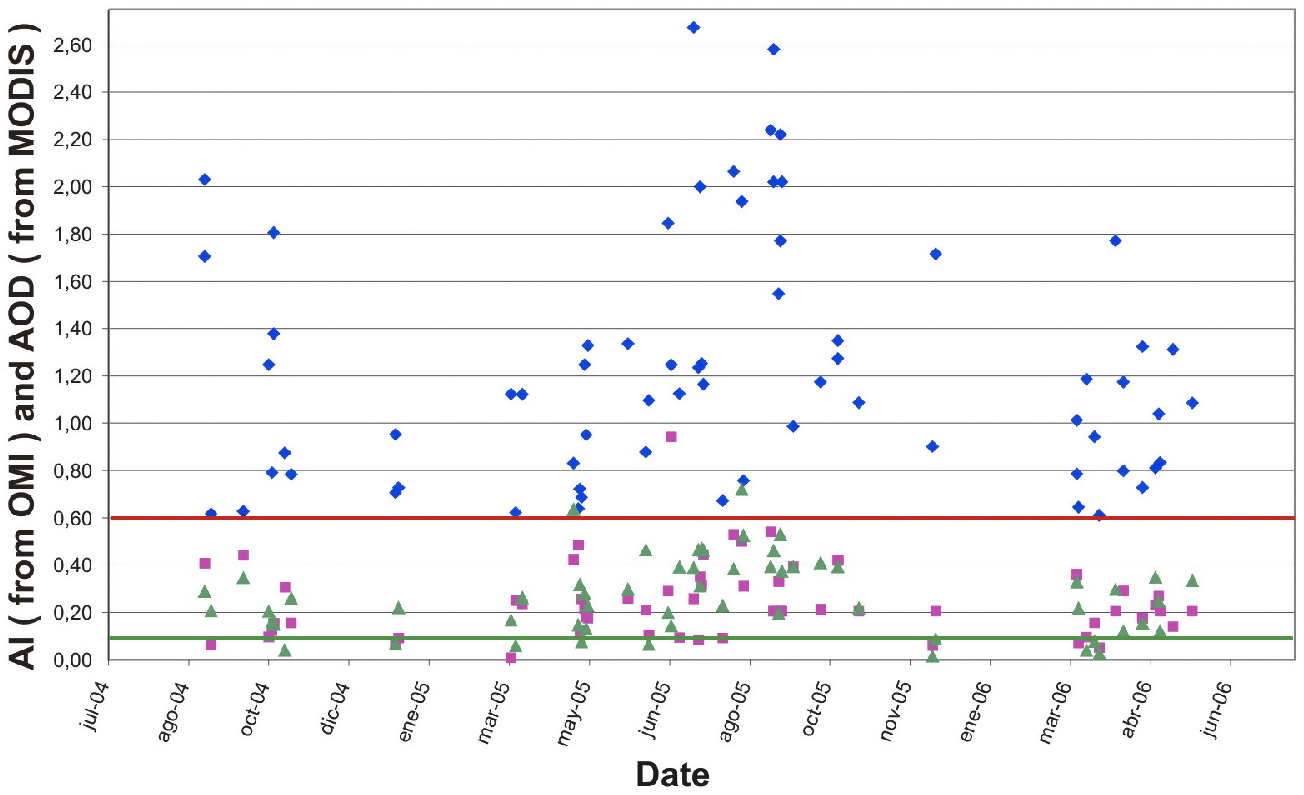}
\caption{AI from OMI (blue rhombus) and AOD from MODIS (pink squares correspond to  Terra values and 
green triangles correspond to  Aqua
data) under dusty conditions detected by the INM from 2004 to 2006.}
\label{fig_limits}
\end{figure*}

To determine the AI and AOD threshold for dusty days we used information provided in a 
collaboration between the Spanish Environment Ministry, 
the Upper Council of Scientific Researches (CSIC) and the National Institute of Meteorology (INM) 
for the study and analysis of the atmospheric pollution produced by airborne aerosols in Spain. They provide us with the 
days in which {\itshape calima} (dust intrusion) occurred in the Canary islands. As we 
later demonstrate in the plots, in our opinion, these 
events happened lower altitudes than those of the 
observatories so they do not influence the measurements of atmospheric extinction. In Fig. \ref{fig_limits} we 
show AOD for Terra and Aqua and AI for OMI. AI values are larger than 0.6, and 
most of AOD data points for Terra and Aqua fall 
above a limit greater than 0.10. This limit is consistent with that provided by Romero \& Cuevas (2002). Therefore, AI$>$0.6
(0.1 units smaller than provided by Siher et al. 2004) and AOD$>$0.1
 will be the thresholds for dusty episodes.

\section {Results II: comparison of AI provided by OMI and AOD provided by MODIS with KV from the CAMC}

In this section we analyse the correlation between the aerosol index (AI) and the aerosol optical 
depth (AOD) provided by 
the satellites and the atmospheric extinction coefficient (KV) measured by {\it in situ} techniques (CAMC).

\subsection{AI from OMI vs KV}

\begin{figure*}
\includegraphics[scale=1]{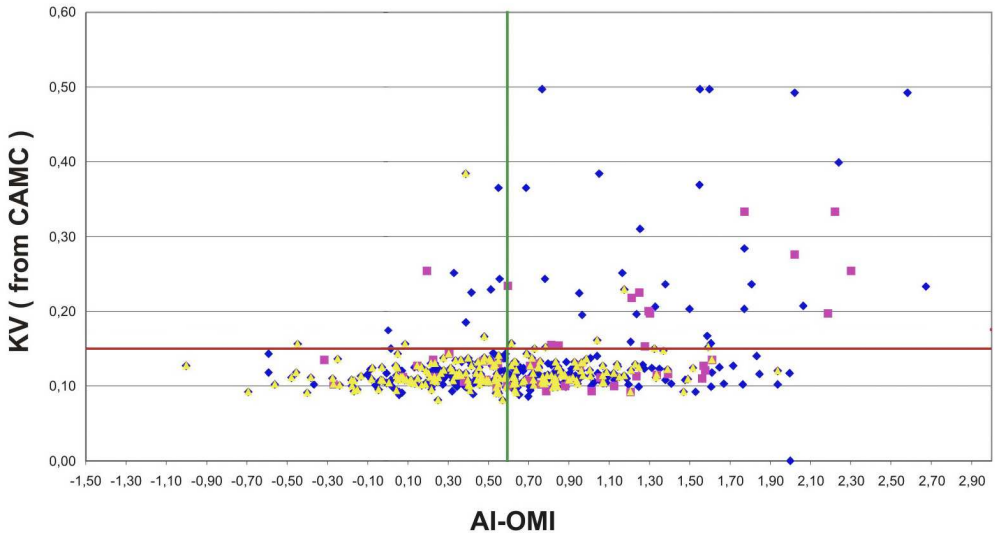}
\caption{Correlation between AI provided by  Aura/OMI and KV from the 
CAMC. Pink points indicate dust episodes  of \textit{calima} retrieved by the INM, and yellow points
correspond to cloud presence. Red and green lines indicate the KV and AI limits for dusty nights respectively.  We can see that most
of the points indicating the presence of absorbing aerosols (AI larger than 0.6) but non atmospheric extinction 
(KV smaller than 0.15 mag/airmass) correspond to the presence of clouds and \textit{calima} below the level of the Observatory.}
\label{fig_corrAI-KV}
\end{figure*}

In this subsection we argue about the correlation between atmospheric extinction in $V$ (551nm) and the aerosol 
index measured for OMI in the same channels of TOMS (331 and 360 nm) represented in Figure \ref{fig_corrAI-KV}. In order to 
determine whether there exists a correlation 
between both parameters we also consider the presence or otherwise of clouds and the situation for days with 
\textit{calima}. 

In Fig. \ref{fig_corrAI-KV} we observe an interesting situation  not revealed in the same graph for the aerosol 
EP/TOMS index   (Fig. \ref{fig_TOMSAI-KV}). Once we put the atmospheric extinction (KV) threshold 
(red line) and the limit for absorbing aerosols (green
line), the situation reveals four quadrants. 
In the first (the upper right) we have a situation with absorbing aerosol and values of atmospheric extinction 
over the limit of photometric dusty nights. In the second quadrant (upper left) there are no points 
that fall inside; this is very 
important 
because it means that non-absorbing aerosols do not influence the extinction above this threshold.
In the third quadrant (bottom left) there are still non-absorbing
aerosols but for clear nights, and this means that this type of sulphate and/or marine aerosol does not 
influence the extinction as we 
expect. In the last quadrant (bottom right) absorbing aerosols are again seen
 in the presence of low extinction values. An explanation 
could be the presence of weakly absorbent particles, such
 as carbonaceous grains or clouds, or a more complex situation with a mixture of cloudy and 
aerosol scenarios, or the presence of absorbing aerosols below the level of the observatory.
The lack of correlation indicates that AI from OMI should not be used for atmospheric extinction characterization at the Roque de los
Muchachos Observatory.
 
We added pink points as  dust episodes  of \textit{calima} retrieved at the ground level, yellow being those that
correspond to cloud presence (reflectivity at 331 nm greater than 15\%). This plot is 
very informative because from the pink points we can see that these \textit{calima} episodes
 are mainly below the threshold of 
dusty nights, this means that they do not  affect the measurements of extinction because the \textit{calima} is at a lower 
level with  respect to the altitude of the observing site,  such dusty events reaching the Roque and being detected
only on some occasions.
We also note that the clouds are almost all below this limit.

This means that values above the threshold of dusty nights are dust absorbing particles at the height of the observatory. 
We see that in this zone of the plot there are almost no clouds and some \textit{calima}
 events that reach the astronomical site. 
Thanks to strong convective motions, they may be driven along the orographic contours of the caldera that borders Roque de 
los Muchachos. In fact, the spatial resolution of OMI ranges from 13 km $\times$ 24 km to 24 km $\times$
 48 km and  can contain part of 
the caldera.

\subsection{AOD from Terra/MODIS vs KV}

The MODIS Aerosol Product monitors the ambient aerosol optical depth over the oceans globally and over a 
portion of the continents.
The aerosol size distribution is also derived over the oceans, and the aerosol type is derived over 
the continents. Therefore,
MODIS data can help us to understand the physical parameters of aerosols affecting the Canarian Observatories.
Daily Level 2 data are produced at 10 km $\times$ 10 km spatial resolution. Aerosols are one of the 
greatest sources of uncertainty in climate modelling. They vary in time and in space, and can lead to variations in cloud 
microphysics, which could impact on cloud radiative properties and on climate. The MODIS aerosol product 
is used to study aerosol 
climatology, sources and sinks of specific aerosol types (e.g.\ sulphates and biomass-burning aerosols), 
interaction of aerosols 
with clouds, and atmospheric corrections regarding remotely sensed land surface reflectance. Above 
land, dynamic aerosol 
models will be derived from ground-based sky measurements and used in the net retrieval process. Over the 
ocean, three parameters 
that describe aerosol loading and size distribution will be retrieved. There are two necessary pre-assumptions 
in the inversion of MODIS data: the first concerns the structure of the size distribution in its entirety and the second one
log--normal modes that will be needed to describe the volume--size distribution: a single mode to describe the 
accumulation mode particles (radius $<0.5$~$\mu$m) and a single coarse mode to describe dust 
and/or salt particles (radius $>1.0$~$\mu$m). The aerosol parameters we therefore expect to retrieve from
the aerosol Level 2 product are 
the ratio between the two modes, the spectral optical thickness and the mean particle size.

The quality control of these products will be based on comparison with ground and climatology stations.

The parameters that we have used from Level 2 in our work are:
latitude and longitude as geolocation values, and AOD (aerosol optical thickness at 0.55~$\mu$m 
for both Ocean [best] and Land [corrected] in a valid range from 0 to 3), aerosol type land that contains 0 = mixed, 1 = dust, 
2 = sulphate, 
3 = smoke, 4 = heavy absorbing smoke, and cloud fraction land and ocean in percentage.

\begin{figure*}
\includegraphics[scale=1]{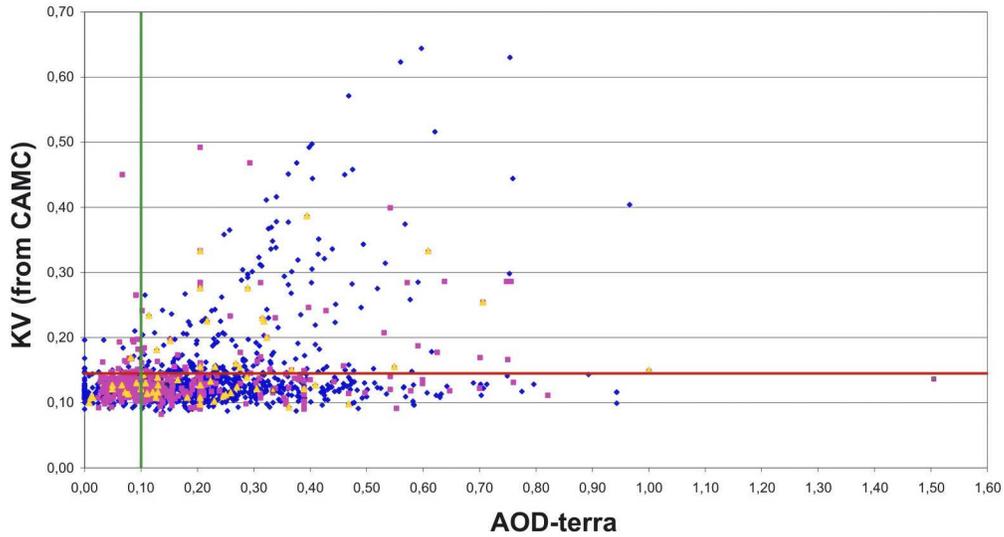}
\caption{Correlation between AOD provided by Terra/MODIS and the KV measured by the CAMC, distinguishing among 
the terrestrial (mixed dust, smoke, sulphate---blue points),
 marine (pink points) aerosols and the presence 
of dust events over the Canary islands coming from Africa and collected at ground level (yellow points). The red and 
green lines 
indicate the KV and AOD limits  respectively for dusty nights.}
\label{fig_corrAODT-KV}
\end{figure*}

In Fig. \ref{fig_corrAODT-KV} we show AOD measured by MODIS on board  Terra against KV.

In this plot we find two tails of data within a certain dispersion  of values, and once more the 
majority of points that fall under the threshold of KV smaller than 0.15. The flatter tail groups together 
most of the AOD values lower than 0.2, i.e. non-absorbing  ($<0.1$) or weakly absorbing  aerosols
(marine particles, clouds above ocean 
and mixed scenarios with salt, sulphate particles and clouds).

In  Fig. \ref{fig_corrAODT-KV} we distinguish among terrestrial aerosols (composed of a mixture of dust and smoke  sulphates)
 and marine
particles marked respectively by blue and pink points and we distinguish the presence 
of dust events coming from Africa (yellow points) using the data provided by INM at ground level. 

We see that pink values are situated in the tails of the plots and are present mainly at lower KV and AOD values, where
 we expect  sulphate sea-salt aerosols and CCN; only in some 
cases are they detected at higher AOD with small KV values, perhaps because these aerosols and clouds are below the
level of the observatory. The second tail covers two quadrants corresponding to
small 
AOD but large KV due to the presence of  weakly absorbent aerosols at the level of the Observatory or to
 a mixture of particles and clouds,  and to large AOD and large KV, mostly terrestrial aerosols at the level 
of the observatory.
 The \textit{calima} (yellow points) is also below the KV threshold, perhaps because of
 their closeness to ground level, except 
for some cases in which it can reach the level of the observatory.

\subsection{AOD from AQUA vs KV}

\begin{figure*}
\includegraphics[scale=1]{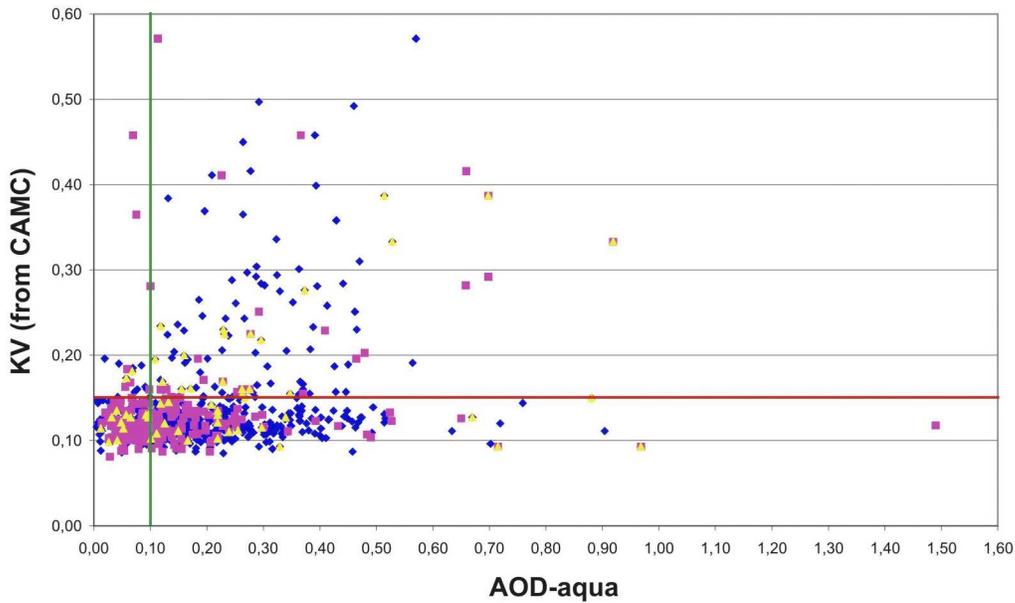}

\caption{Correlation between AOD provided by  Aqua/MODIS and the KV measured by the CAMC, distinguishing 
among the terrestrial (mixed dust, smoke, sulphate---blue points),
 marine (pink points) aerosols and the presence 
of dust events over the Canary islands coming from Africa and collected at ground level (yellow points). 
The red and green lines 
indicate the KV and AOD limits  respectively for dusty nights.}
\label{fig_corrAODA-KV}
\end{figure*}

We now perform the same analysis for the MODIS instrument on board the
Aqua satellite that retrieves 
the same 
aerosol Level 2 data. In the Fig. \ref{fig_corrAODA-KV} we show all the Aqua data in order to see the correlation
 between AOD 
and KV. In this case also there is evidence of the two tails with a wide dispersion in the data, here, most of the
 points being 
below the KV threshold.

We plot terrestrial (blue) and marine (pink) aerosols, the  yellow points indicating days with \textit{calima}. 
The comments 
are very similar to those made previously, i.e. the marine aerosols are more clustered
at small AOD values in a lower tail, whereas 
the terrestrial values are somewhat uniformly distributed with some points at large KV indicating absorbing aerosols or 
clouds at the
level of the observatory.
Most of points fall below the threshold for dusty nights just as the majority of \textit{calima} 
days. The interesting fact is that only the episodes of dust with large KV follow a 
good linear relation with the AOD.

\section{Summary and Outlook}

 From this study we may draw the conclusions listed below.

We have compared the AI measurements provided by the TOMS on board Earth Probe satellite (Level 3) 
with the atmospheric 
extinction coefficient provided by the CAMC  at the ORM (2400 m above MSL). The main causes of the lack of
 correlation
between both parameters are:

\begin{itemize}
\item	The TOMS Level 3 data considered in this paper have a resolution of 1$^{\circ}\times 1.25^{\circ}$, 
so the AI is averaged over areas whose size covers the entire islands of La Palma and Tenerife. 
High resolution Level 2 data should be used for a better fit (IFOV of 35 km $\times$ 35 km).
\item	The TOMS is very sensitive to the presence of highly reflective clouds because it uses channels 
centred on the UV to measure AI. Moreover, AI incorporates absorbing particles in ranges that do not affect 
atmospheric transparency in the visible range.
\item TOMS measurements are retrieved at local noon while CAMC values are averaged over night hours.
 \end{itemize}

For this reason the EP/TOMS database is not useful for the characterization of the presence 
of dust above either 
the Canarian astronomical observatories  (2400 m above mean sea level) or for other high mountain sites 
(Mauna Kea and
San Pedro M\'artir).

We have explored  the use of other detectors on board different satellites that operate in bands of 
astronomical interest (the visible and NIR) and with  better spatial resolution than TOMS. The selected parameters were the
aerosol index provided by Aura/OMI---with visible and ultraviolet channels and with a spatial resolution from 
13 km $\times$ 24 km
to 24 km $\times$
48 km---and the aerosol optical depth (AOD) provided by Terra (from 2000) and Aqua (from 2002) 
in MODIS---with its 36 spectral bands, from 0.47 to 14.24 $\mu$m, including two new channels at 0.405 and 0.550 $\mu$m, 
with a spatial resolution of 10 km $\times$ 10 km.
In order ot obtain the best spatial, spectral, radiometric and temporal resolutions, we have decided to work only 
with Level 2 data that have the same resolution as the IFOV satellite.

We conclude that the OMI instrument   detects aerosol presence with more 
precision than TOMS and does not detect
non-absorbing particles with high atmospheric extinction 
values (larger than 0.15 mag/airmass). This fact coincides with  expectations because non-absorbing aerosols 
such as sulphates or marine aerosols do not give high extinction values (threshold greater than 0.15 mag/airmass). 
We can see that most of points fall at lower extinction values below the threshold for dusty nights, 
suggesting presence of non-absorbing or weakly absorbing (e.g. carbonaceous) aerosols. 
   
In order to obtain the limits for dusty episodes on the AOD scale, we have checked the \textit{calima} days
from the records of the Instituto Nacional de Meteorolog\'{\i}a de Canarias Occidental, following NAAPS, ICOd/DREAM or SKIRON models,  
and we obtain a threshold near AOD$>$0.1 units and AI$>$0.6 for dusty episodes.

We study where the \textit{calima} events and cloud presence fall in the plot of 
correlation between atmospheric extinction and AI. 
Dust episodes measured at  ground level are mainly below the threshold for dusty nights on the 
atmospheric extinction scale (KV$<$0.15 mag/airmass), meaning that  the presence of \textit{calima} affects low altitudes, and that
only in a few 
cases does it reach the Roque. We also see that all clouds detected for their high reflectivity (greater than 15\%)
are below the threshold for dusty nights; in only  two cases are they above 0.15 unit in mag/airmass and in such 
cases they do not correspond to \textit{calima} events. 

Study of Terra/MODIS data has shown that a great number of points fall in 
a range below 0.40 units for AOD and 0.15 mag/airmass 
for KV, and that two tails are evident: the first one has high AOD values for low KV and the second has high AOD and
high KV, showing a large linear correlation among both parameters. 
Is important to underline that  there are only a few points with low AOD values and high KV (data are decontaminated 
of cloud presence). 
Chlorides and marine aerosols can be well 
identified and normally do not affect the KV and correspond to AOD $<$ 0.2, so  this means that marine and 
sulphate aerosols 
are not absorbent as we expect. 
We can see that the most of points corresponding to dust (\textit{calima}) events fall below 
the threshold of 0.15 mag/airmass  because they are detected near the surface; only in certain cases can  they reach the 
level of the observatory, carried 
by wind or convective motions, providing AOD larger than 0.1 units and KV larger than 0.15. 
These measurements correspond mostly to terrestrial aerosols.

The study of the Aqua data produces results almost identical to those of Terra.
In this case there are also 
values at high AOD and low KV and values at low AOD and KV. This wide dispersion explains the lack of
correlation between both parameters. The most populated tail falls below the threshold of 0.15 unit in mag/airmass. 

Marine aerosols are more clustered at low AOD values ($<$ 0.20) and low KV ($<$0.15), whereas   terrestrial 
aerosols fall 
in the 
  AOD$>$0.2 zone. Dusty (\textit{calima}) days  correspond mostly to the presence of terrestrial aerosols and
 are present near  ground level (inside the area of 10 km $\times$ 10 km), so they appear at low KV. Only 
in some cases do they reach higher 
altitudes and become detectable from the astronomical observations (large KV); this is the only case that presents 
a linear correlation
between both parameters.  We therefore also need {\it in situ}  data to distinguish between both situations. We must explore other 
clues for the vertical aerosol drainage analysis (winds, humidity, etc.).

At present, the AI and AOD values provided by the NASA satellites are not  useful  for 
aerosol site characterization, and {\it in situ} 
data are required to study drainage behaviour, in particular at those astronomical sites with abrupt orography 
(ORM, Mauna Kea or San Pedro M\'artir). Spatial resolution of the order of the observatory area
 will be required in these cases.

Moreover,
in order to obain much better spatial resolution we are now exploring the use of SEVIRI-MSG2 
(1.4 km $\times$ 1.4 km) (December 2005--12) for Europe and Africa, 
 and  in the future, ATLID(LIDAR)-EARTHPROBE (with a horizontal sampling interval smaller than 100 m) 
   (2012--15) for global coverage. The CALIPSO satellite will be used for measuring the vertical structure (drainage) 
   and properties of 
  aerosols (the horizontal
and vertical spatial resolution are 5 km and 60 m respectively).

Ground measurements will be complemented by LIDAR data (INTA) of 30 m resolution (NASA MPL-NET-AERONET) and by the IAC
airborne particle counter (from Pacific Scientific Instruments) installed at the ORM in February 2007 
(with six channels: 0.3, 0.5, 
1, 3, 5, 10~$\mu$m)
and by the INM  
Multifilter Rotating Shadowband Radiometer (MFRSR) programmed to be installed at the ORM in the near future (consisting of 
six narrow passbands between 414 nm and 936 nm) that will provide the size, density and vertical distribution of
the aerosols.

Tests with AERONET {\it in situ} data during daytime could be very interesting to make a comparison with remote sensing 
satellite data.

\section*{Acknowledgments}

We express our deepest thanks to the TOMS, OMI and MODIS groups from NASA Goddard Space Flight Center for 
aerosol index and aerosol
optical depth measurements, and to the Carlsberg 
Meridian Circle  of the Isaac Newton Group on La Palma for the coefficient extinction data. Our 
acknowledgments go to
the Main directorate of Quality and Environmental Evaluation of the Environment Ministry, 
the Superior Council of Scientific Researches (CSIC) and the National Institute of Meteorology (INM) of 
the Environment Ministry  
for the information on the atmospheric pollution produced by  airborne aerosols in Spain. 
This study is part of the site characterization work developed by the Sky Quality Group of the IAC and
 has been carried out 
within the framework of the European Project OPTICON and under   Proposal FP6 for Site Selection for the European ELT.
Many thanks are also due to the anonymous referee, whose comments and suggestions helped us to improve the article.

\section*{Appendix I}

\section*{Format of satellite data}

In this appendix we describe the format of the data for the different satellites 
used in this work   and 
we indicate the urls from which  to retrieve the datasets.

\subsection*{I.1  N7/TOMS (1984-1993)}

Level 2 data are available at http://daac.gsfc.nasa.gov/data/dataset/TOMS/Level\_2/N7/ and
were ordered with ftp-push. NASA loaded the data onto our ftp url. These are raw data in hdf5 format that 
must be seen with HDFview and afterwards processed with computer programs. 
Level 3 aerosol data are available at ftp://toms.gsfc.nasa.gov/pub/nimbus7/data/aerosol
and are already available in ftp format. These are daily average data with a resolution of 
1.25$^{\circ}$ in longitude  and 1$^{\circ}$ in latitude. These data are in ASCII format with 288 bins in longitude, 
centred on 179.375 W to 179.35 E, every bin has a 1.25$^{\circ}$ step. In latitude there are 180 bins centred on 
89.5 S to 89.5 N with a step of 1$^{\circ}$. The values are in 3 digit groups, 
missing data being tagged 999 and the other numbers being multiplied by 10.

\subsection*{I.2  EP/TOMS (1996-2005)} 

 Level 2 data are available at http://daac.gsfc.nasa.gov/data/dataset/TOMS/Level\_2/EP/.
The format and the data processing is the same as for Nimbus7. Level 2 data are produced with a 
spatial resolution of 39 km $\times$ 39 km at nadir. Level 3 aerosol data are immediately available at
 ftp://toms.gsfc.nasa.gov/pub/eptoms/data/aerosol.

It is important to underline that the aerosol monthly average datasets are computed using only positive values 
(i.e. absorbing aerosol indices) of the aerosol index for each month. Values of zero are used in the averaging whenever
the aerosol index is negative. The final monthly average datasets contain aerosol index values greater 
than or equal to 0.7.

\subsection*{I.3 Aura/OMI (2004-NOW)}
 
Level 2 data with a spatial resolution of 13 $\times$ 24 km at  nadir are available at 
http://daac.gsfc.nasa.gov/data/dataset/OMI/Level2/\\OMT03/.
They can be ordered via ftp-push in the same way as the others. To obtain information about effective cloud pressure and 
fraction data go to http://daac.gsfc.nasa.gov/data/dataset/OMI/Level2/\\OMCLDRR/.

OMTO3 provides aerosol index, total column ozone and aerosol optical thickness, as well as ancillary information 
produced from the 
TOMS Version 8 algorithm applied to OMI global mode measurements. In the global mode each file contains a single orbit of 
data covering a width of 2600 km.
Compared to TOMS, OMI's smaller field of view results in a larger ``sea glint''  per unit field of view and a  correspondingly 
larger error in derived ozone under these conditions. The OMTO3 aerosol index is not valid for solar zenith angles greater 
than 60$^{\circ}$. Because the OMI solar zenith angles are typically higher than the solar zenith angles for TOMS at the 
same latitude, the OMI AI becomes invalid at somewhat lower latitudes than TOMS. This may show a cross-track dependence 
in the OMI AI and is not corrected by the radiance measurement 
adjustments (error in the AI up to 4\%). Compared to TOMS, the OMTO3 Aerosol Index is 0.5 NVALUE high. Users of the aerosol 
index are advised to make this correction for consistency with the TOMS data record.
For users not interested in the detailed information provided in OMTO3 dataset several 
gridded products are being developed. 
Initially, DAAC will grid OMTO3 data in a format identical to that used for TOMS (1$^{\circ}$ 
$\times$ 1.25$^{\circ}$ lat/long) 
and will make it available through the TOMS website. However, to take advantage of the higher spatial resolution of the 
OMI products DAAC intends  to produce higher resolution gridded products for all OMI datasets, including OMTO3.

\subsection*{I.4 Terra/MODIS (2000-NOW)} 

 Level 2 data are available at 
http://daac.gsfc.nasa.gov/data/dataset/MODIS/\\02\_Atmosphere/01\_Level\_2/01\_Aerosol\_Product/index.html.
We can order them with ftp-push. These are raw data in hdf5 format and so we must 
follow a similar procedure to the one we use 
with TOMS. Daily Level 2 (MOD04-Terra) aerosol data are produced at the spatial resolution of 10 $\times$ 10 km at nadir. 
We can also retrieve data  via ftp (they are already available on the web) through the url 
ftp://g0dps01u.ecs.nasa.gov/MODIS\_terra\_Atmosphere/\\MOD04\_L2.004.

Level 3 data are available at http://disc.sci.gsfc.nasa.gov/data/dataset/MODIS/\\02\_Atmosphere/02\_Level\_3/.
These are different atmospheric data, daily, weekly and monthly averaged in a global 1$^{\circ}$ $\times$ 1$^{\circ}$ grid. 
The method to order this dataset is ftp-push: The aerosol information is stored in MOD08\_D3, MOD08\_E3 and MOD08\_M3.
Note that in MODIS there is no aerosol index but only the aerosol optical thickness.

\subsection*{I.5  Aqua/MODIS (2002-NOW)} 
The same as for Terra/MODIS. The urls are: 
http://daac.gsfc.nasa.gov/data/dataset/MODIS-Aqua/02\_Atmosphere/01\_Level\_2/01\_Aerosol\_Product/\\index.html 
(the ftp address ftp://g0dps01u.ecs.nasa.gov/MODIS\_\\Aqua\_Atmosphere/MYD04\_L2.004) for Level 2 data and 
http://disc.sci.gsfc.nasa.gov/data/\\dataset/MODIS-Aqua/02\_Atmosphere/02\_Level\_3/ for the Level 3 data.

\section*{Appendix II}
                    
AEMET Agencia Estatal de Meteorolog\'{\i}a \\
AEROCE   Atmosphere/Ocean Chemistry Experiment \\
AERONET   Aerosol Robotic NETwork \\
AI   Aerosol Index \\
AOD   Aerosol Optical Depth \\
AQUA   Earth Observing System Post Meridian (PM) \\
ASCII   American Standard Code for Information Interchange \\
ATLID   ATmospheric LIDAR \\
AURA   Earth Observing System Chemistry mission \\
CALIPSO Cloud-Aerosol Lidar and Infrared Pathfinder Satellite \\
CAMC   Carlsberg Automatic Meridian Circle Telescope \\
CCN   Cloud Condensation Nuclei \\
CSIC   Consejo Superior de Investigaciones Cient\'{\i}ficas \\
ELT   Extremely Large Telescopes \\
ENVISAT   ENVIronmental SATellite \\
EP   Earth Probe \\
ERS   European Remote Sensing Satellites \\
ESA    European Space Agency   \\                  
EUMETSAT   EUropean METeo SATellite \\
ENVISAT  ENVironmental SATellite\\
FP6 Sixth Framework Programme \\
FTP   File Transfer Protocol \\
GMT   Greenwich Mean Time \\
GOME   Global Ozone Monitoring Experiment \\
GSFC   Goddard Space Flight Center \\
HDF   Hierarchical Data Format \\
HTTP   Hyper Text Transfer Protocol \\
IAC   Instituto de Astrof\'{\i}sica de Canarias \\
IAU   International Astronomical Union \\
ICoD/DREAM  Dust Loading model forecast from Insular Coastal Dynamics \\	 
IFOV   Istantaneous Field of View \\
INM   Instituto Nacional de Meteorolog\'{\i}a \\
INTA   Instituto Nacional de T\'ecnica Aerospacial \\
IP   Internet Protocol \\
L2   Level 2 data \\
L3   Level 3 data \\
KV   Atmospheric extinction coefficient in V-band \\
LIDAR   Light Detection And Ranging \\
MFRSR   MultiFilter Rotating Shadowband Radiometer \\
MML   Maritime Mixing Layer \\
MODIS   Moderate Resolution Imaging Spectroradiometer	 \\
MPLNET    MicroPulse Lidar NETwork \\
MSG   Meteosat Second Generation \\
MSL   Mean Sea Level \\
N7   Nimbus-7 \\
NAAPS   Navy Aerosol Analysis and Prediction System \\
NASA   National Aeronautics and Space Administration \\
NEODC   NERC Earth Observation Data Center \\
NERC   Natural Environment Research Council \\
NIR   Near Infra Red  \\
NRT   Near Real Time \\
NUV   Near Ultra Violet \\
O3   Ozone \\
OAI   Iza\~na Atmospheric Observatory \\
OMI   Ozone Monitoring Instrument \\
OMTO3   OMI Total Column Ozone \\
OPTICON   Optical Infrared Coordination Network for Astronomy \\
ORM   Roque de Los Muchachos Observatory  \\
OT   Teide Observatory \\
SCIAMACHY   Scanning Imaging Absorption SpectroMeter Atmospheric Chartography \\
SEVIRI   Spinning Enhanced Visible and InfraRed Imager \\
SKIRON   Weather Forecasting Model operated by University of Athens \\
TERRA   Earth Observing System Anti Meridian (AM) \\
TL   Troposphere Layer \\
TOMS   Total Ozone Mapping Spectrometer \\
UK   United Kingdom \\
UV   Ultra Violet \\


\label{lastpage}
\end{document}